\documentclass[letterpaper,twoside,12pt]{article}
\usepackage{amssymb}
\usepackage{amsmath}
\usepackage{latexsym}
\usepackage{verbatim}
\usepackage{graphicx}
\usepackage{epsfig}
\usepackage{stmaryrd}
\usepackage{rotating,graphicx}
\usepackage[english]{babel}
\usepackage{setspace}
\usepackage[english]{babel}
\usepackage[utf8]{inputenc}
\usepackage[T1]{fontenc}
\usepackage{lmodern}
\usepackage{pstricks}

\begin{document}

%%%%%%%%%%%%%%%%%%%%%%%%%%%%%%%%%%%%%%%%%%%%%%%%%%%%%%%%%%%%%%
%%%%%%%%%%%%%%%%%%%%%%%%%%%%%%%%%%%%%%%%%%%%%%%%%%%%%%%%%%%%%%
\newcommand{\bra}[1]{\langle #1|}
\newcommand{\ket}[1]{|#1\rangle}
\newcommand{\braket}[2]{\langle #1|#2\rangle}
%%%%%%%%%%%%%%%%%%%%%%%%%%%%%%%%%%%%%%%%%%%%%%%%%%%%%%%%%%%%%%
%%%%%%%%%%%%%%%%%%%%%%%%%%%%%%%%%%%%%%%%%%%%%%%%%%%%%%%%%%%%%%

\begin{Large}
\begin{center}
\textbf{Less Decoherence and More Coherence in Quantum Gravity, Inflationary Cosmology and Elsewhere}\\%[-5cm]
\end{center}
\end{Large}

\begin{center}
\begin{large}
Elias Okon${}^1$ and Daniel Sudarsky${}^2$\\[.3cm]
\end{large}
${}^1$\textit{Instituto de Investigaciones Filos\'oficas, Universidad Nacional Aut\'onoma de M\'exico, Mexico City, Mexico.}\\
${}^2$\textit{Instituto de Ciencias Nucleares, Universidad Nacional Aut\'onoma de M\'exico, Mexico City, Mexico.}
\end{center}

\noindent \textbf{Abstract:} 
In \cite{Crull} it is argued that, in order to confront outstanding problems in cosmology and quantum gravity, interpretational aspects of quantum theory can by bypassed because decoherence is able to resolve them. As a result, \cite{Crull} concludes that our focus on conceptual and interpretational issues, while dealing with such matters in \cite{Oko.Sud:14}, is avoidable and even pernicious. Here we will defend our position by showing in detail why decoherence does not help in the resolution of foundational questions in quantum mechanics, such as the measurement problem or the emergence of classicality.
%%%%%%%%%%%%%%%%%%%%%%%%%%%%%%%%%%%%%%%%%%%%%%%%%%%%%%%%%%%%%%
\section{Introduction}
\label{Int}
%%%%%%%%%%%%%%%%%%%%%%%%%%%%%%%%%%%%%%%%%%%%%%%%%%%%%%%%%%%%%%
\onehalfspacing
%%%%%%%%%%%%%%%%%%%%%%%%%%%%%%%%%%%%%%%%%%%%%%%%%%%%%%%%%%%%%%
Since its inception, more than 90 years ago, quantum theory has been a source of heated debates in relation to interpretational and conceptual matters. The prominent exchanges between Einstein and Bohr are excellent examples in that regard. An avoidance of such issues is often justified by the enormous success the theory has enjoyed in applications, ranging from particle physics to condensed matter. However, as people like John S. Bell showed \cite{Bell1}, such a pragmatic attitude is not always acceptable. 

In \cite{Oko.Sud:14} we argue that the standard interpretation of quantum mechanics is inadequate in cosmological contexts because it crucially depends on the existence of observers external to the studied system (or on an artificial quantum/classical cut). We claim that if the system in question is the whole universe, such observers are nowhere to be found, so we conclude that, in order to legitimately apply quantum mechanics in such contexts, an observer independent interpretation of the theory is required. Nowadays, there are a number of versions of quantum theory, such as Bohmian mechanics \cite{Gold} and objective collapse models \cite{GRW}, with the needed characteristic. In \cite{Oko.Sud:14} we focus on the latter to display an array of benefits that such theories offer regarding the resolution of long-standing problems in cosmology and quantum gravity. In particular, we explore applications of objective collapse theories to the origin of seeds of cosmic structure, the problem of time in quantum gravity and the information loss paradox.

In a recent paper entitled ``Less Interpretation and More Decoherence in Quantum Gravity and Inflationary Cosmology,'' E. M. Crull argues that our focus on conceptual and interpretational issues in \cite{Oko.Sud:14} is unjustified because quantum decoherence ``addresses precisely those aspects of the measurement problem many believe require resolution before going onwards, and for resolving new issues within relativistic applications of the theory'' \cite[p. 1020]{Crull}. Crull also maintains that decoherence is all is needed in order to explain the quantum-to-classical transition and the emergence of classicality. It seems, then, that it is not that she believes that the issues we raise, regarding, e.g., the necessity to use an observer independent version of quantum mechanics in cosmological settings, are not critical, but that she thinks that decoherence is capable of resolving them. 

The objective of this work is twofold. We will defend our position by pointing out some of the most glaring shortcomings in \cite{Crull}. However, before doing so, we will explain in detail why decoherence fails as a tool to resolve foundational questions in quantum mechanics, such as the measurement problem and the quantum-to-classical transition. The plan for the manuscript is as follows: in section \ref{Deco} we will critically evaluate the relevance of decoherence in foundational discussions, in section \ref{Crull} we will directly address the criticism of Crull and in section \ref{Con} we will present our conclusions.

%%%%%%%%%%%%%%%%%%%%%%%%%%%%%%%%%%%%%%%%%%%%%%%%%%%%%%%%%%%%%%
\section{Decoherence}
\label{Deco}
%%%%%%%%%%%%%%%%%%%%%%%%%%%%%%%%%%%%%%%%%%%%%%%%%%%%%%%%%%%%%%
In recent years, decoherence has become an extremely popular subject, both in applied and fundamental physics. In the former, decoherence is a subject of enormous relevance, as the effect is one of the main concerns in the practical development of quantum computers as well as in other interesting experimental proposals. In the latter, it has often been promoted as the solution to long-standing foundational problems. In particular, decoherence is deemed to solve (at least important parts of) the measurement problem and to explain the quantum-to-classical transition. The idea is that decoherence, which is a straightforward consequence of a purely unitary evolution, is able to effectively explain or bring about the collapse of the wave function, which is, of course, at the root of the measurement problem. Similarly, decoherence is said to dynamically single out a preferred basis, which is supposed to coincide with the classical one, thus explaining the emergence of classicality. The objective of this section is to carefully show that, contrary to widespread believe, decoherence does not help in the resolution of these foundational questions. The main reason being, as we will see below in detail, that in order for decoherence to accomplish what it is supposed to, one needs to assume the very thing that is to be achieved.

Quantum decoherence arises from i) the fact that the inevitable interactions between a system and its environment typically lead to entanglement between the two and ii) the fact that the states of the environment entangled with the different states of the system, usually become, very quickly and to a very good approximation, orthogonal among themselves. These facts are often described in a more ``poetic'' language which, sometimes, is not straightforward to interpret formally. For example, Crull defines decoherence as follows:
\begin{quotation}
\noindent [I]t is a dynamical process whereby a system's phase relations in particular bases become decohered or randomized by commuting [sic] with external (environmental) degrees of freedom.\footnote{Probably she meant to say ``coupling'' instead of ``commuting''.} \cite[p. 1021]{Crull}
\end{quotation}
However, it is rather unclear what lies behind the idea that the system's phase relations become ``decohered or randomized.'' Although one might have some intuitive feeling about things becoming random when they are too complicated, the fact is that the unitary evolution provided by the Schr\"odinger equation contains absolutely nothing random. Moreover, no clear, unambiguous and universally accepted definition of coherence (which is supposed to get lost in the process) is available, For example, in \cite{Sch}, which has lately become the standard reference on the subject, a state is said to be in a \emph{coherent superposition} when each of the component states $|\psi_n\rangle$ is simultaneously present in the state $|\Psi\rangle = \sum_n c_n|\psi_n\rangle$. It is hard to see what the adjective of ``coherent'' adds to the standard notion of a superposition and what is the part that gets lost through the purely unitary interaction at play. It could be that what is really meant is that when the state of a composite systems is entangled, then each component does not possess a well-defined state. However, if that is all there is to the loss of coherence, then one must conclude that any entangled system, such as an EPR pair, is not coherent. In fact, below we will describe in detail what is (probably) meant by the language in the quote above. All we want to point out here is that such expressions are often repeated without an adequate understanding of the fact that they are only figures of speech to be unpacked in formal terms, and that such unpacking is much more subtle and complex than is usually thought.

This way of talking about decoherence is even more pervasive, and more confusing, when the consequences of the process are discussed. For example, \cite[p. 5]{Sch} states:
\begin{quotation}
\noindent [T]he coupling to the environment now defines the observable physical properties of the system. At the same time, quantum coherence, a measure for the ``quantumness'' of the system, is delocalized into the entangled system–environment state, which effectively removes it from our observation.
\end{quotation}
in the classic \cite[p. 5]{Zur} Zurek writes:
\begin{quotation}
\noindent [S]ystems usually regarded as classical suffer (or benefit) from the natural loss of quantum coherence, which
``leaks out'' into the environment... The resulting ``decoherence'' cannot be ignored when one addresses the problem of the reduction of the quantum mechanical wave packet: Decoherence imposes, in effect, the required ``embargo'' on the potential outcomes by allowing the observer to maintain records of alternatives but to be aware of only one of the branches -- one of the ``decoherent histories'' in the nomenclature of Murray Gell-Mann and James Hartle...
\end{quotation}
and Crull gives us:
\begin{quotation}
\noindent Decoherence of a system will suppress to extraordinary degree interference terms in the decohered basis, such that further interactions will practically always “see” the system in an eigenstate of the basis or bases most affected by decoherence (that is, with respect to system degrees of freedom that commute most rapidly and efficiently with environmental degrees of freedom). \cite[p. 1021]{Crull}
\end{quotation}
We believe that all these words in quotations only obscure the meaning of these phrases and, more importantly, complicate a conceptual appraisal of their validity. The point, as we said above, is that one must not forget that these expressions are only figures of speech and that, in order to fully evaluate them, they have to be unpacked in formal terms. The problem, as we will see below in detail, is that, in doing so, one finds that decoherence in fact does not accomplish what it is supposed to. 

In order to start such unpacking, we note that decoherence is supposed to solve the measurement problem and to explain the emergence of classicality by accomplishing two things (see \cite[p. 6]{Sch}):
\begin{enumerate}
\item Suppression of macroscopic interference.
\item Selection of a preferred basis.
\end{enumerate}
In words of Zurek:
\begin{quotation}
\noindent Decoherence destroys superpositions. The environment induces, in effect, a superselection rule that prevents certain superpositions from being observed. Only states that survive this process can become classical. \cite[p. 21]{Zur}
\end{quotation}
Therefore, in order to evaluate the ability of decoherence to solve foundational issues, one must check if decoherence accomplishes these two facts. Below we will show in detail why decoherence in fact does \emph{not} explain the suppression of macroscopic interference nor selects a preferred basis. Therefore, decoherence is unable to solve the measurement problem and to explain the emergence of classicality. In order to fully back up these claims (sections \ref{si} and \ref{pb} below), we will start with some necessary preliminary material.

%%%%%%%%%%%%%%%%%%%%%%%%%%%%%
\subsection{The standard formalism and the measurement problem}
%%%%%%%%%%%%%%%%%%%%%%%%%%%%%
In order to properly discuss the merits of decoherence as a tool to solve foundational problems in quantum mechanics, it is very useful to start by considering first the standard interpretation and, in particular, by being explicit about what it consists of. The standard interpretation of quantum mechanics, for which we mean something along the lines of the version of Dirac \cite{Dirac}, von Neumann \cite{von} or what one usually finds in contemporary textbooks, can be summarized as follows:
\begin{itemize}
 \item The physical states of quantum systems are represented, at all times, by unit vectors in an appropriate Hilbert space.
 \item The physical properties of the system are represented by Hermitian operators.
 \item The time evolution of the system is governed by the, fully deterministic, Schr\"odinger equation.
 \item The connexion between the mathematical formalism and predictions is given by the Born rule, which allows one to compute the possible results for any experiment, along with the probabilities associated with each possible result.
 \item Finally, there is the projection or collapse postulate that states that, after a measurement, the state of the system instantaneously jumps to the eigenstate of the measured property with the eigenvalue corresponding to the measured value.
\end{itemize}
As it is well known, if the state of the system, before a measurement, is an eigenstate of what is going to be measured, then Born's rule predicts that, with certainty, the result will be the associated eigenvalue. From this, together with some type of a realist criterion, along the lines of the famous EPR sufficient condition for an element of reality \cite{EPR}, one can conclude what is usually called the Eigenvector-Eigenvalue (EE) rule, which states: 
\begin{itemize}
\item A state possesses the value $\lambda$ of a property represented by operator the $O$ if and only if that state is an eigenvector of $O$ with eigenvalue $\lambda$.
\end{itemize}

At any rate, what is clear is that, within the standard interpretation, the only way to connect the mathematical representations (i.e., vectors and operators) with predictions, is trough the Born rule (and possibly through the EE rule as well). That is, the only way to extract a claim regarding what to expect with respect to the physical properties of a quantum system, from the fact that its quantum state is such and such, is via the Born rule (an possibly the EE rule as well). In this regard, the Born and EE rules play the role of \emph{physical interpretations} of the mathematical formalism of standard quantum mechanics. That is, they constitute the dictionary that translates between the math and the world that the physics tries to describe. 

But what happens if one decides to drop the Born rule and the collapse postulate, as is proposed, e.g., within the decoherence program in order to deal with the conceptual difficulties of quantum theory? Then one must necessarily substitute this interpretation of the mathematics of the formalism for something else. If one does not do that, then we do not have, properly speaking, a physical theory. That is, the link between the formalism and physical predictions is lost. Moreover, by considering a purely unitary formalism, i.e., by dropping the collapse postulate (and, with it, the Born rule), one must be very careful not to unwittingly, and of course invalidly, use the standard interpretation of the formalism in terms of collapses and probabilities. We will see below that much of the confusion regarding the reach of decoherence as a tool to resolve foundational questions arises from ignoring these aspects and making this type of mistakes.

The formalism described above is, of course, the most successful empirical theory ever constructed. The problem, however, is that, in spite of such an amazing predictive success, the formalism is not fully satisfactory -- the culprit being the notorious measurement problem. But what is the measurement problem? Crull is right in pointing out that the measurement problem means different things to different people and that, in fact, it sometimes consists of a group of interrelated issues and not only one specific problem. She is also right in pointing out that there is a lot of confusion in the literature regarding its true nature, import and consequences. Unfortunately, we believe that the discussion of the problem in \cite{Crull} only deepens the misunderstanding. 

The measurement problem is, of course, a problem of a theoretical framework. Therefore, in order to state the problem clearly, it is crucial to first specify in detail the theoretical framework in question. However, given the proliferation of views regarding quantum mechanics, it is no wonder that there exists an accompanying proliferation of ways to state the problem. And given that many of these different ways to present the quantum formalism are not always compatible among themselves, it is, again, no wonder that the many different ways to present the measurement problem are also not compatible among themselves. For example, in terms of the standard interpretation discussed above, the problem manifests as the fact that the formalism crucially depends on the notion of measurement, but that such notion is never precisely defined within the theory.\footnote{In section 2.1 of \cite{Crull} Crull briefly discusses the role measurements play in quantum mechanics, but by doing so she only contributes to a long tradition of fallacious statements regarding the issue. To start off, she claims that entanglement arises even in the absence of interaction, which is simply wrong. Then she tries to define a measurement entity as something capable of gaining information about some system, such that the information can later be gathered. Such definition is, of course, so vague that it is practically useless. Moreover, it is circular because in order to gather such information at a later time one, presumably, needs to somehow measure it!} %
The result is a formalism that, under some circumstances, is vague. To other characterizations of quantum mechanics correspond different formulations of the measurement problem: in Bohr’s formulation, the problem manifests as an ambiguity regarding where the classical-quantum cut should be drawn, and, in a theory with purely unitary evolution (like the one considered within the decoherence program) it manifests as a mismatch between experience and some predictions of the theory (e.g., Schrödinger cat states).

Among all the formulations of the measurement problem present in the literature, Crull decides to use the one in \cite{Sch}. According to Schlosshauer's book, the measurement problem is composed of three distinct parts:
\begin{enumerate}
\item \emph{The problem of the preferred basis}. What singles out the preferred physical quantities in nature -- e.g., why are physical systems usually observed to be in definite positions rather than in superpositions of positions?
\item \emph{The problem of the nonobservability of interference}. Why is it so difficult to observe quantum interference effects, especially on macroscopic scales?
\item \emph{The problem of outcomes}. Why do measurements have outcomes at all, and what selects a particular outcome among the different possibilities described by the quantum probability distribution?
\end{enumerate}
Schlosshauer argues that decoherence resolves the first two components and the first part of the third one (i.e., why measurements have outcomes at all). He also claims that it is debatable whether it solves the last bit of the third one since it is linked to the choice of interpretation. In any case, he claims that this last element of the problem is only of ``philosophical'' relevance and not important for practical purposes. 

At any rate, given our previous discussion regarding the nature of the measurement problem, it is clear that this formulation of the problem is defective. That is because it does not, in any way, identify the precise problem it wants to display within a detailed and specific theoretical framework. Instead, particularly in parts 1 and 2, it simply states some empirical facts. It is clear that, what Schlosshauer actually has in mind, is a formalism with a purely unitary evolution in which, as we said above, the measurement problem manifests as a mismatch between experience and some predictions of the theory. However, by not being explicit about this in his presentation, he obscures some of the crucial issues regarding the conceptual analysis needed to evaluate his claims. In particular, by not stating precisely what is the concrete quantum formalism he endorses, he allows for some ``wiggle room'' at the moment of \emph{physically interpreting} the formal results obtained by decoherence.

Therefore, in order to avoid confusion, we will be explicit about the quantum formalism in play. It consists of the first three elements of the standard formalism described above. That is: physical states are represented by vectors, properties by Hermitian operators and states \emph{always} evolve according to the Schrödinger equation. Note, that by dropping the collapse postulate and the Born rule, as we explained above, one looses the standard physical interpretation of the mathematical formalism (this will play a crucial role later on when we evaluate the ability of decoherence to solve the measurement problem). 

With this quantum formalism in place, it is easy to show the discrepancy between its predictions and experience. In this regards, Zurek in \cite[p. 4]{Zur} states:
\begin{quotation}
\noindent [A]t the root of our unease with quantum theory is the clash between the principle of superposition -- the basic tenet of the theory 
reflected in the linearity of [the Schrödinger Equation] -- and everyday classical reality in which this principle appears to be violated.
\end{quotation}
More formally, consider a quantum system $S$ with Hilbert space $H$ and consider an observable $A$ with eigenvectors
 $\{\ket {a_i}_S\}$. Consider also a measuring device $M$, with ready state $\ket {r}_M$, which 
 interacts with the system in the following way
\begin{equation}
 \ket {a_i}_S \ket {r}_M \xrightarrow{Schrödinger} \ket {a_i}_S \ket {a_i}_M
\end{equation}
for all $i$, where $\ket {a_i}_M$ is the (macroscopic) state of $M$ which indicates that the result of the measurement is $a_i$. That is, $M$ is a good measuring device. Then, by the linearity of the Schrödinger equation, it is clear that
\begin{equation}
 \left( \sum_i c_i \ket {a_i}_S \right) \ket {r}_M \xrightarrow{Schrödinger} \sum_i c_i \ket {a_i}_S \ket {a_i}_M .
\end{equation}
Therefore, the prediction of the theory is that the final state of the measurement is a superposition of different (macroscopic) states for $M$, which is not what we observe in the lab. Note however that Schlosshauer's way of stating the measurement problem, i.e., the particular list of facts he chooses regarding the incompatibility between theory and experience, is suspiciously convenient for decoherence in light of what it is supposed to achieve (i.e., suppression of macroscopic interference and selection of a preferred basis). That is, from among all the discrepancies between a purely unitary formalism and experience, he chooses to highlight precisely those that exactly correspond to what he believes decoherence explains. For example, he highlights the absence of macroscopic interference but it is more than an absence of interference that is observed but not predicted. In fact, in order to claim that interference does not occur we would need to perform the appropriate experiments, but the discrepancy between experience and the predictions of a purely unitary quantum theory is there even if we do not perform such experiments. What we actually see is the absence of superpositions of $M$ in certain bases.

Below we will evaluate in detail the decoherence process and its alleged ability to solve foundational questions. However, before doing so, it is important to say a few words regarding the usage of density matrices within quantum theory.
%%%%%%%%%%%%%%%%%%%%%%%%%%%%%
\subsection{Density matrices}
%%%%%%%%%%%%%%%%%%%%%%%%%%%%%
A density matrix is an operator that encodes information about a quantum mechanical system. However, density matrices are used for different purposes and, in different situations, they contain different amounts and kinds of information regarding the quantum mechanical system (or systems) in question (see e.g., \cite{Sch,despagnat}).

To begin with, we point out that, according to the standard interpretation of quantum mechanics, all closed or isolated quantum systems possess, at all times, well-defined quantum states. These states are refereed to as \emph{pure} and are usually represented by vectors on an appropriate Hilbert space. However, as is well known, pure states can also be represented by density matrices. In particular, if the pure state of a system is given by $| \psi \rangle$, then one can define its associated density matrix by $\rho_p = | \psi \rangle \langle \psi|$, which satisfies $\rho_p^2=\rho_p$ and $Tr(\rho_p)=1$.\footnote{For any operator $A$, its trace is defined by $Tr(A)=\sum_i \langle\phi_i|A|\phi_i\rangle$ with $\{\phi_i\}$ any basis of the Hilbert space in question.} It is clear that, in this pure state case, the density matrix $\rho_p$ contains exactly the same information that the original state vector $| \psi \rangle$. In fact given a density matrix satisfying $\rho_p^2=\rho_p$ and $Tr(\rho_p)=1$, one can recover the corresponding unity norm vector of the Hilbert space (up to an irrelevant complex phase). Moreover, it is easy to show that one can use $\rho_p$ to compute expectation values via $\langle O\rangle=Tr(\rho_p O)$. Note that this identification of the (mathematical notion) trace with the (physical concept) expectation value is legitimate only if one assumes the Born rule and the collapse postulate. That is, such mathematical object can be interpreted in terms of an expectation value, i.e., in terms of what one expects the average of a large number of measurements on an ensemble of identically prepared systems to be, only if one assumes that during each measurement, one finds as a result one of the eigenvalues of $O$, with probabilities given by the Born rule. This clarifications might seem trivial at this point but will be crucial latter when we evaluate the claims regarding the alleged achievements of decoherence.

Density matrices can also be used to describe ensembles of identical systems in which not all members of the ensemble are prepared in the same state. Instead, the states of the different members are distributed among a set of states $\{|\chi_i\rangle\}$, with respective frequencies $p_i$. To such situation, one assigns the density matrix $\rho_m=\sum_i p_i | \chi_i \rangle \langle \chi_i|$, which satisfies $Tr(\rho_m)=1$. It is clear that, in the $\{|\chi_i\rangle\}$ basis, the constructed matrix is diagonal, with diagonal elements given by the frequencies $p_i$. The motivation behind this definition for the density matrix comes from the fact that, in terms of $\rho_m$, the expectation values can, again, be calculated via $\langle O\rangle=Tr(\rho_m O)$. Note however that, in this case, the calculation of the expectation value depends on two types of probabilities: the quantum ones, as in the previous case, but also on the classical or \emph{epistemic} ones, associated with a lack of knowledge of the exact state of any individual member of the ensemble. Therefore, in this case, the density matrix contains only statistical information regarding the possible state of any particular member. A closely related application of density matrices involves the description of situations in which one wants to study a closed quantum system, but one does not have full information regarding its pure state and only knows that the state is one of the members of $\{|\chi_i\rangle\}$ with respective probabilities $p_i$. In such case, one again assigns the density matrix $\rho_m=\sum_i p_i | \chi_i \rangle \langle \chi_i|$ and calculates expectation values via $\langle O\rangle=Tr(\rho_m O)$. Both of these cases are usually refereed to as \emph{mixed} states or \emph{proper} mixtures.

%Density matrices can also be used to describe situations in which one wants to study a closed quantum system, but one does not have full information regarding its pure state and only knows that the state is one of the members of $\{|\chi_i\rangle\}$ with respective probabilities $p_i$. To such situation one assigns the density matrix $\rho_m=\sum_i p_i | \chi_i \rangle \langle \chi_i|$, which satisfies $Tr(\rho_m)=1$. It is clear that, in the $\{|\chi_i\rangle\}$ basis, the constructed matrix is diagonal, with diagonal elements given by the probabilities $p_i$. The motivation behind this definition for the density matrix comes from the fact that, in terms of $\rho_m$, the expectation values can, again, be calculated via $\langle O\rangle=Tr(\rho_m O)$. Note however that, in this case, the calculation of the expectation value depends on two types of probabilities: the quantum ones, as in the previous case, but also on the classical or \emph{epistemic} ones, associated with a lack of knowledge of the exact state of the system under study. Therefore, in this case, the density matrix contains only statistical information regarding the possible states of the system. A closely related application of density matrices involves the description of ensembles with different members possessing different states. In such case, the $p_i$ introduced above refer to the \emph{frequency} with which the $\{|\chi_i\rangle\}$'s occur in the ensemble. Both of these cases are usually refereed to as \emph{mixed} states or \emph{proper} mixtures.

Before moving on, it is important to point out an ambiguity regarding mixed states. The problem is that, sometimes, \emph{different} mixed states are assigned the \emph{same} density matrix. As a result, if one is given a density matrix of a mixed state, one does not automatically know which are the possible states for the system(s) described. To illustrate the point, consider two ensembles of electrons, one with half of the electrons with spin-up and half with spin-down along $z$ and the other with half with spin-up and half with spin-down along $y$. Given that
\begin{equation}
\label{e3}
\frac{1}{ 2}\left\{ |+ \rangle_{z} \langle+|_{z}+|- \rangle_{z} \langle-|_{z} \right\} = \frac{1}{ 2}\left\{ |+ \rangle_{y} \langle+|_{y}+|- \rangle_{y} \langle-|_{y} \right\} , 
\end{equation}
it is clear that both ensembles, which clearly correspond to different physical situations, will be associated with the same density matrix.

Finally, there is the application of density matrices for the description of a subsystem of a (closed or isolated) quantum system. Suppose, then, that a system $S$ is composed of two subsystems $A$ and $B$. Of course, in general, the partition of $S$ into subsystems $A$ and $B$ is arbitrary. Now, given that $S$ is an isolated quantum system, it invariably possesses a well-defined quantum state. However, if such state is an entangled state between $A$ and $B$, it is impossible to assign well-defined states to such subsystems. That is, the quantum formalism entails that, in such situations, the subsystems simply do not possess well-defined states. There is, however, a way to encode \emph{some} of the information regarding a subsystem in a density matrix. In order to do so, one first constructs the pure density matrix associated with the whole system $S$ and then takes the partial trace of such matrix over the rest of the system. That is, if $| \Psi \rangle$ is the state of $S$, one defines $\rho_A = Tr_B(\rho)$ with $\rho = | \Psi \rangle \langle \Psi|$. If, on one hand, the state of $S$ is separable, the subsystem $A$ possesses a pure state and $\rho_A$ is just the corresponding pure density matrix. If, on the other hand, the state of $S$ is entangled, $A$ does not possess a well-defined state so $\rho_A$ cannot represent its state. Now, if one considers observables of $S$ of the form $O_s= O_A \otimes \mathbb{I}$, then it is easy to show that $\langle O_s\rangle=Tr(\rho_A O_A)$. Therefore, as we said above, $\rho_A$ does not contain any information regarding the actual physical state of the subsystem in question -- such subsystem simply does not possess one. Instead, it contains information regarding expectation values of \emph{measurements} that could be performed on such a subsystem. Density matrices of this type are called \emph{reduced} density matrices or \emph{improper} mixtures, and satisfy $Tr(\rho_A)=1$.

At this point we would like to make a few comments that are crucial in order to evaluate the claims regarding the capacity of decoherence to solve foundational problems. These comments have to do with similarities and differences regarding mixed states and reduced density matrices. Regarding similarities, it is clear that, mathematically speaking, they are identical. That is, they are both generically represented by matrices with trace equal to one. As a result, entangled subsystems and ensembles are often described with matrices of identical form. Regarding differences, it is central to keep in mind that the physical situations described by mixtures (or proper mixtures) and reduced density matrices (or improper mixtures) are extremely different. In the first case, the systems described always possess well-defined quantum states, even though these might not be known to us. In the second case, if the subsystem one wants to describe is entangled, it simply does not possess a well-defined state. As a result, $\rho_A$ cannot be considered to represent the state of the system, so it becomes merely a mathematical tool to encode information regarding the behavior of $A$ in \emph{measurement} situations. That is, it is only an instrument useful to make predictions regarding the results one expects to find while performing measurements on the system in question, assuming, of course, both the Born rule and the projection postulate.

Now we are in position to evaluate if, as advertised, decoherence is in deed able to suppress macroscopic interference and choose a preferred basis. We will discuss each of these in turn.

%%%%%%%%%%%%%%%%%%%%%%%%%%%%%
\subsection{Suppression of interference}
\label{si}
%%%%%%%%%%%%%%%%%%%%%%%%%%%%%
The starting point of the claim that decoherence explains the absence of macroscopic interference, is the fact that every object inevitably interacts with its environment and, furthermore, that through such interaction the environment ``obtains certain information'' regarding the system. Moreover, it is claimed that the states of the environment that correspond to different states of the system are almost orthogonal. The environment, then, is said to ``continually measure'' the system and it is stressed that such ``measurement'' does not require a human observer of any sort. 

Now let us see how all this works more formally. Suppose we have a quantum system whose initial state is some superposition $\frac{1}{\sqrt{2}} \left( |\psi_1 \rangle + |\psi_2 \rangle \right)$ and we let it interact with its environment, with initial state $|E_0 \rangle$. As a result of what we said above, the evolution will look as follows
\begin{equation}
\label{sE}
\frac{1}{\sqrt{2}} \left( |\psi_1 \rangle + |\psi_2 \rangle \right) |E_0 \rangle \xrightarrow{Schrödinger} \frac{1}{\sqrt{2}} \left( |\psi_1 \rangle |E_1 \rangle+ |\psi_2 \rangle |E_2 \rangle \right) ,
\end{equation}
with $\langle E_1|E_2 \rangle \approx 0$. Next it is argued that, since we are interested in the system, and not the environment, we can trace over the degrees of freedom of the environment and construct a reduced density matrix for the system, $\rho_S$, which contains a complete description of its measurement statistics. The result is
\begin{eqnarray}
\rho_S &=& \frac{1}{2} \left\{ |\psi_1\rangle\langle\psi_1|+|\psi_2\rangle\langle\psi_2|+|\psi_1\rangle\langle\psi_2|\langle E_2|E_1 \rangle+|\psi_2\rangle\langle\psi_1|\langle E_1|E_2 \rangle\right\} \nonumber \\
&\approx& \frac{1}{2} \left\{|\psi_1\rangle\langle\psi_1|+|\psi_2\rangle\langle\psi_2|\right\} .
\end{eqnarray}
What we obtain, then, is a reduced density matrix, formally identical to that of a mixed state in which the system in question is in either the state $|\psi_1\rangle$ or the state $|\psi_2\rangle$, each with probability of $1/2$. Does this mean that this is then the case for the system we are considering? That is, that the system in question is suddenly, no longer in the state on the LHS of Eq. (\ref{sE}), in which the system does not even have a well-defined state, but instead, the system is, definitely, either on the state $|\psi_1\rangle$ or on the state $|\psi_2\rangle$? Of course not. As we explained before, the fact that a reduced density matrix an a mixed state can have the same form does not mean that they represent the same physical situations and, in this case, it is crystal clear that they do not. The mere fact of deciding to ignore the degrees of freedom of the environment of course cannot have any physical impact on the state of the system. At any rate, the claim defended by decoherence enthusiasts is often more sophisticated than what we have discussing so far. The claim they maintain is not that, under the circumstances we have been discussing, the system is on either $|\psi_1\rangle$ or $|\psi_2\rangle$, but that, \emph{for all practical purposes}, it will behave as if it where. Let us spell out the argument.

The state of the system is of course given by the LHS of Eq. (\ref{sE}). However, it is noted that only measurements that include both the system and the environment will be able to corroborate it and that, in practice, it is impossible to keep track of all of the environmental degrees of freedom. As a result, it is argued that $\rho_S$ is, \emph{for all practical purposes} the tool to use in order to make predictions regarding all possible measurements to be carried out on the system. And since $\rho_S$ is identical to a mixed state, the results of all these possible measurements are going to be identical to those of measurements performed on a mixed state. That is, \emph{for all practical purposes} the system will behave as a mixture. Decoherence, then, is said to lead to \emph{effectively non-unitary dynamics} for the system, which explains the absence of interference between the components of the superposition. In this regards, Zurek writes:
\begin{quotation}
\noindent The key advantage of [a diagonal reduced density matrix] is that its coefficients may be interpreted as classical probabilities. The [reduced] density matrix... can be used to describe the alternative states of a composite spin-detector system that has classical correlations. Von Neuman's process 1 serves a similar purpose to Bohr's ``border'' even though process 1 leaves all the alternatives in place. When the off-diagonal terms are absent, one can nevertheless safely maintain that the apparatus, as well as the system, is each separately in a definite but unknown state, and that the correlation between them still exists in the preferred basis defined by the states appearing on the diagonal. \cite[p. 8]{Zur}
\end{quotation}

All this sounds very attractive, the problem is that the above argument is fallacious. In order to understand why, we need to remember a couple of things. First, that the quantum formalism at play in decoherence is purely unitary. Therefore, in order to make empirical predictions, it cannot make use of the standard interpretation of the mathematical apparatus in terms of the probabilities dictated by the Born rule. Second, that the possibility to interpret a reduced density matrix as a tool to make predictions, i.e., the possibility to read its entries as probabilities, crucially depends on assuming the Born rule. Therefore, the interpretation of the reduced density matrix needed for decoherence to work as claimed is not available to decoherence. That is, in order for decoherence alone to solve the measurement problem, it would need to presuppose exactly what it is trying to explain. In fact, within a purely unitary formalism, such as the one considered by decoherence, not only one cannot interpret a reduced density matrix in terms of probabilities but, since no substitution for the standard interpretation is given, no predictions can be extracted at all. We conclude that, contrary to widespread believe, decoherence by itself is not able to explain the absence of macroscopic interference nor, as a result, to solve the measurement problem.

%%%%%%%%%%%%%%%%%%%%%%%%%%%%%
\subsubsection{A simple example}
\label{ex}
%%%%%%%%%%%%%%%%%%%%%%%%%%%%%
If one takes seriously the claims regarding the achievements of decoherence, then Shrödinger’s cat, usually considered a paradoxical situation, is in fact fully and satisfactorily resolved when one decides to trace over the state of the atomic nucleus whose decay would trigger the release of poisonous gas. That is because, doing so, leads to a full decoherence of the density matrix characterizing the poor cat. We, after all, are only concerned about the cat. According to the program of decoherence, in its application to foundations, as we are only interested in the cat and not the atom, we are justified not only in tracing over the atom's degrees of freedom but in interpreting the diagonal nature of the reduced density matrix as indicating that, \emph{for all practical purposes}, the cat is either dead or alive. However, in order to be consistent, we must insist that a similar position be taken in all instances where the corresponding situation arises. That is, we must insist that whenever one obtains a density matrix that is diagonal as the result of tracing over certain degrees of freedom, one should be able to adopt the analogous position, i.e., that the system of interest is, \emph{for all practical purposes}, in one of the states that are represented in that diagonal. Let us now test whether this is indeed a tenable position. In order to do that, we will consider a standard EPR-B type situation. 

Assume that a spin-0 particle at rest decays into two identical spin-1/2 particles in an angular momentum conserving process. Let us call $x$ the axis aligned with the momenta of the emerging spin-1/2 particles. We characterize the two-particle state that results from the decay in terms of the $z$ polarization states of the two Hilbert spaces of the individual particles. As the angular momentum of the two particle state must vanish, it follows that the state must be
\begin{equation}
|\phi \rangle= \frac{1}{\sqrt2} \left\{ |+ \rangle^{(1)}_{z} |- \rangle^{(2)}_{z} + |-\rangle^{(1)}_{z} |+ \rangle^{(2)}_{z} \right\} ,
\end{equation}
and the density matrix for the system is given by $\rho= |\phi\rangle\langle\phi|$. Suppose now that we decide that we are not interested in one of the particles (call it $1$), and thus we regard it as an ``environment'' for the system of interest (particle $2$). We therefore, decide to trace over its degrees of freedom to obtain the following reduced density matrix for particle $2$
\begin{equation}
\rho^{(2)} = Tr_1 (\rho)= \frac{1}{ 2}\left\{ |+ \rangle^{(2)}_{z} \langle+|^{(2)}_{z}+|- \rangle^{(2)}_{z} \langle-|^{(2)}_{z} \right\} ,
\end{equation}
which clearly is diagonal. Therefore, what we have is a completely decohered density matrix so, according to the attitude described above, particle $2$ must be considered as having a definite value, of either $+1/2 $ or $-1/2$, for its spin along the $z$ axis. However, it is clear that taking such position in not viable.

We can start by noting that the fact that the state $|\phi\rangle$ is symmetric with respect to rotations around the $x$ axis implies that we could have written the density matrix using instead the $y$ polarization states of the two Hilbert spaces of the individual particles (see Eq. (\ref{e3})). That is,
\begin{equation}\label{DM}
\rho^{(2)} = \frac{1}{ 2} \left\{ |+ \rangle^{(2)}_{y} \langle+|^{(2)}_{y}+|-
\rangle^{(2)}_{y} \langle-|^{(2)}_{y} \right\} , 
\end{equation}
leading this time to the conclusion that the particle must be considered as having its spin along the $y$ axis defined to be either $+1/2 $ or $-1/2$. So which one is it?, does particle $2$ have a well defined value of its spin along $z$ or $y$? Clearly the approach does not lead to a coherent position.

Furthermore, given Aspect's experiments confirming the violation of Bell's inequalities \cite{Aspect1,Aspect2,Aspect3}, it follows that one cannot assume that particle $2$ has a definite (even if unknown) value for its spin before a measurement takes place. Clearly, such position would lead to the problematic conclusions highlighted by Bell \cite{Bell}.

It could be pointed out that in order to verify the violation of Bell's inequalities, one needs to compare results from both particles and that, at such point, it is no longer true that we are ignoring the environment (i.e., particle $1$). That is, at such point, the \emph{for all practical purposes} clause is violated because it is no longer true that we, in effect, only have access to part of the system. That may be true but it does not solve the problem for decoherence. That is because, besides the issue of the basis ambiguity mentioned above, it is at this point that if one wants to describe particle $2$ with a reduced density matrix, and use it to make predictions regarding what one expects to see, one need to invoke the Born rule. That is, in order to interpret the diagonal elements of the reduced density matrix as \emph{probabilities}, one need to assume that, upon measurement, one will find the particle on an eigenstate of what one measures, with probabilities given by Born's rule.

%%%%%%%%%%%%%%%%%%%%%%%%%%%%%
\subsection{Preferred basis}
\label{pb}
%%%%%%%%%%%%%%%%%%%%%%%%%%%%%
As we explained before in detail, a standard way to state the measurement problem is as a mismatch between experience and some predictions of standard unitary quantum mechanics. More concretely, the problem corresponds to a discrepancy between the prediction of the widespread presence of macroscopic superpositions and the fact that observers always end up with determinate measurement results. An alternative way to present the measurement problem is as the fact that, even though the standard formulation of quantum mechanics depends crucially on the notion of measurement, such notion is never formally defined within the theory. Then, in order to apply the formalism, one needs to know, by means external to quantum mechanics, what constitutes a measurement, when a measurement is taking place, and what it is that one is measuring.

From all of the above, an important component of the measurement problem, usually refereed to as the \emph{basis problem}, can be isolated. In the first case, it corresponds to the fact that not only the predictions of the formalism deviate form experience but, since it treats all bases on an equal footing, it does not even single out a particular basis in which such determinate results are supposed to occur (this in fact corresponds to the second point in Schlosshauer's formulation). For the second formulation of the measurement problem above, the basis problem presents itself as the inability of standard quantum mechanics to ascertain in advance, and without information external to the formalism itself, what it is that is going to be measured in any particular measurement situation. It is clear then that any solution of the measurement problem needs also to address the basis problem.\footnote{In \cite{Crull}, the basis problem is associated with the following question: ``Given the statistical improbability of always observing bases that are classical, why should such preferences for them appear in nature?'' We find the decision to state the problem in terms of a statistical improbability quite curious since one does not expect the observed basis to be chosen at random.} 

Let us illustrate this with a simple example. Consider a quantum system $S$ (with a 2D Hilbert space) whose state, in the basis associated to observable $A$, is given by the following superposition
\begin{equation}
\ket \psi_S = \alpha \ket {a_1}_S + \beta \ket {a_2}_S .
\end{equation}
Consider also, again, a measurement apparatus $M$ with ready state given by $\ket {r}_M$ and which interacts with the system in the following way
\begin{equation}
\ket \psi_S \ket {r}_M = \left ( \alpha \ket {a_1}_S + \beta \ket {a_2}_S \right ) \ket {r}_M \xrightarrow{Schrödinger} \alpha \ket {a_1}_S \ket {a_1}_M + \beta \ket {a_2}_S \ket {a_2}_M .
\end{equation}
Of course, one can write the state $\ket \psi_S $ in the basis of observable $B$ instead
\begin{equation}
\ket \psi_S = \gamma \ket {b_1}_S + \delta \ket {b_2}_S ,
\end{equation}
in which case the interaction with the apparatus looks as follows
\begin{equation}
\ket \psi_S \ket {r}_M = \left ( \gamma \ket {b_1}_S + \delta \ket {b_2}_S \right ) \ket {r}_M \xrightarrow{Schrödinger} \gamma \ket {b_1}_S \ket {b_1}_M + \delta \ket {b_2}_S \ket {b_2}_M .
\end{equation}
Now consider the following question: if we perform the experiment in the laboratory, will we observe the final state of the apparatus to be either $\ket {a_1}_M$ or $\ket {a_2}_M$ or either $\ket {b_1}_M$ or $\ket {b_2}_M$ (or neither)? As we explained above, the standard interpretation is unable to answer such a question \emph{unless} external information, to the effect that $M$ actually measures (let's say) $A$, is provided. Given that such information is not contained in the standard fundamental description of the situation (given by the quantum states described above), we conclude that the standard interpretation does not solve the measurement problem. The purely unitary formalism, at least \emph{prima facie}, is also not able to answer because it treats all bases on an equal footing. Decoherence, notwithstanding, is supposed to be able to fix the problem. Let us see how this is supposed to work.

The idea, as described in \cite[p. 73]{Sch} is the following:
\begin{quotation}
\noindent The \emph{preferred states} of the system emerge dynamically as those states that are the least sensitive, or the most \emph{robust}, to the interaction with the environment, in the sense that they become least entangled with the environment in the course of the evolution and are thus most immune to decoherence.
\end{quotation}
More formally, the preferred states are supposed to be the ones that satisfy
\begin{equation}
|\psi_i\rangle|E_0\rangle \xrightarrow{Schrödinger} |\psi_i\rangle|E_i\rangle ,
\end{equation}
with $\langle E_i|E_j \rangle \approx 0$. The idea, then, is that states that are not altered through the interaction with the environment are deemed stable and thus observable. Conversely, states that do change are said to decohere rapidly and to become, as a result, unobservable in practice. The reasoning behind this is that, on the one hand, initial superpositions in the preferred basis will evolve, as in Eq. (\ref{sE}), to entangled states. As a result, their reduced density matrices will be diagonal, so such states are going to be observable. On the other hand, Initial superpositions in other bases will not decohere, in the sense that their reduced density matrices will not be diagonal. In consequence, such states are going to be unobservable in practice. 

In praise of the proposal, Schlosshauer writes:
\begin{quotation}
\noindent The clear merit of the approach of environment-induced superselection to the preferred-basis problem lies in the fact that the preferred basis is not chosen in an \emph{ad hoc} manner so as to simply make our measurement records determinate or to match our experience of which physical quantities are usually perceived as determinate (for example, position). Instead the selection is motivated on physical, observer-free grounds, namely, through the structure of the system–environment interaction Hamiltonian... The appearance of classicality is therefore grounded in the structure of the physical laws governing the system–environment interactions. \cite[p. 85]{Sch}
\end{quotation}
Still, a couple issues come to mind. The first one is whether the rule given to fix the preferred basis is really observer-free, as advertised. The answer, of course, is that it is not because the division of the world into a system and an environment is totally arbitrary. As a result, different decisions as to how to split a system will lead to different preferred bases. One could try to argue that our epistemic limitations as humans determine which degrees of freedom are accessible to us, and thus dictate a particular way so select the environment. However, that is very different from claiming, as in the quote above, that the selection of the basis is observer-free.

The second issue is the fact that the offered explanation of why it is that we have access to one basis, the preferred one, but not others, crucially depends on the alleged suppression of interference achieved by decoherence as long as the reduced density matrix in question is diagonal. However, we already saw that such link between diagonality and observability is fallacious. As a result, the fact that the density matrix in one basis is diagonal does not imply that such basis will be special in any empirically interesting sense. Therefore, even if one concedes the fact that the selection of the basis is observer-free, one still does not have a satisfactory explanation for the preferred basis.

%%%%%%%%%%%%%%%%%%%%%%%%%%%%%
\subsection{Is decoherence helpful to foundations?}
%%%%%%%%%%%%%%%%%%%%%%%%%%%%%
Summing up, decoherence is supposed to achieve two things: the suppression of macroscopic interference and the selection of a preferred basis. With respect to the former, we have seen that the claim that a decohered system behaves, \emph{for all practical purposes}, like a classical mixture, is not warranted. The reason being that in order to show that the decohered system in deed behaves like the mixture, one needs to assume the Born rule. That is, one need to assume that when one measures, one always finds as a result an eigenstate of the measured operator. Therefore, in order to show that one will not find a superposition when one measures, one needs to precisely assume that one will not find a superposition when one measures, defeating the purpose of the whole enterprise.

Regarding the selection of a preferred basis, we have seen first that the process is not as observer-free as advertised because the partition of the world into a system and an environment is of course arbitrary. Furthermore, we have pointed out that the argument that allegedly explains why it is that there is a dynamically emergent preferred basis crucially depends on the ability of decoherence to suppress macroscopic interference. Given that decoherence fails in such a task, the argument in favor of the preferred basis crumbles. We conclude, then, that decoherence does not explain the nonobservability of interference nor the emergence of a preferred basis. As a result, it also does not solve the measurement problem nor explains the emergence of classicality.

The starting point of Crull's criticism of our work is the fact that decoherence solves all the relevant interpretational and conceptual problems we worry about. Then she argues that, since such problems were the motivation behind our decision to consider non-standard interpretations, the invocation of a specific interpretation\footnote{Given that the particular interpretation we consider in \cite{Oko.Sud:14} is fundamentally indeterministic, we find it odd for Crull to claim that the urgency to consider a specific interpretation most often arises from a hesitation to accept that the world is indeterministic.} %
becomes not only unnecessary but burdensome. Here we have shown that decoherence does not address the pressing foundational issues of quantum mechanics. We trust that such state of affairs emphatically vindicates the motivations behind our work. In the next section we will examine in more detail specific criticisms of Crull regarding \cite{Oko.Sud:14}.

%%%%%%%%%%%%%%%%%%%%%%%%%%%%%%%%%%%%%%%%%%%%%%%%%%%%%%%%%%%%%%
\section{Response to Crull}
\label{Crull}
%%%%%%%%%%%%%%%%%%%%%%%%%%%%%%%%%%%%%%%%%%%%%%%%%%%%%%%%%%%%%%

Section 4 in \cite{Crull} opens with some remarks regarding our motivations, described in \cite{Oko.Sud:14}, for focusing on objective collapse models, as well as with comments about our take on the measurement problem.\footnote{Apparently, Crull finds our brief review of basic features of objective collapse models in \cite{Oko.Sud:14}, which she takes to be a definition of such models, unsatisfactory: ``one might argue that the way in which [Okon and Sudarsky] define objective collapse theories introduces as many black boxes as it purports to explain''. It is unclear what is it that she finds in need of further explanation. Evidently, if one is looking for a completely viable collapse model compatible with relativistic quantum field theory, one will not find it in our work, nor elsewhere, since such a theory is still very much under construction. Therefore, one should not compare it directly with finished proposals, such as ``decoherence'' or the ``Consistent Histories'' approach. That is, one cannot compare directly programs under development, such as quantum gravity proposals, with well established theories such as general relativity, and demand the former to be as precisely formulated at this stage as is the latter. On the other hand, one must recognize the potential of the former to deal with evident shortcomings of the latter (i.e., the incompatibility of GR with quantum theory). At any rate, the literature on objective collapse models is of course large and of excellent quality (see e.g., \cite{GRW} and references therein).} %

Then, Crull discusses our application of objective collapse models to the three problems considered in \cite{Oko.Sud:14} and she describes what, according to her, decoherence reveals in each of those situations. Below we will directly address such an analysis, but before doing so, we will present a quick overview of what we defend in \cite{Oko.Sud:14}.

%%%%%%%%%%%%%%%%%%%%%%%%%%%%%
\subsection{``Benefits of Objective Collapse Models for Cosmology and Quantum Gravity'' in a nutshell}
\label{Ben}
%%%%%%%%%%%%%%%%%%%%%%%%%%%%%

As we mentioned above, the notion of measurement plays a central role in the standard formulation of quantum mechanic; yet, such formalism offers no clear rules to determine which interactions should count as measurements or which subsystems as observers. The problem is that the predictions of the theory crucially depend on how this so-called ``Heisenberg’s cut'' is implemented. Moreover, when the system to be studied is the whole universe, the complications deepen because, in such case, there is nothing outside the system that could play the role of the observer. All this is, in essence, the measurement problem of quantum theory, the resolution of which has motivated the development of various alternative versions and modifications of quantum mechanics. Clearly, in order to apply quantum theory to cosmology, one of these alternatives, one that is observer-independent, is required.

Nowadays there are several proposals for a quantum formalism not fundamentally based on the notion of measurement or on that of an observer external to the system under consideration. They include (with various degrees of success) Bohmian mechanics, Many-world scenarios and several objective collapse models. In \cite{Oko.Sud:14} we focused on the latter in order to highlight their potential for the resolution of some long-standing problems in cosmology and quantum gravity. Objective collapse theories modify the dynamical equation of the standard theory, with the addition of stochastic and non-unitary terms, designed to account, on the basis of a single law, for both the quantum behavior of micro-systems and the absence of superpositions at the macro-level (without ever having to invoke observers or measurements). Early examples of such theories include CSL \cite{Pea:89} and GRW \cite{GRW:86}, and recently even fully relativistic versions have been developed \cite{Tum:06,Bed:11}.

In \cite{Oko.Sud:14} we noted the potential of these theories to resolve three important open issues in cosmology and quantum gravity: the origin of the seeds of cosmic structure, the problem of time in quantum gravity and the black hole information loss paradox. Below we briefly describe how objective collapse models are able to help in their resolution.
\begin{itemize}
 \item The inflationary period of cosmological evolution is supposed to erase all memory of initial conditions, leading to a completely flat, homogeneous and isotropic early universe with quantum fields in fully homogeneous and isotropic quantum states. The seeds of cosmological structure, which bring about the formation of galaxy clusters, galaxies and stars, and thus represent a departure from such symmetry, are then supposed to emerge as a result of quantum fluctuations. The standard account for the formation of such seeds implicitly assumes a transition from a symmetric quantum state to an essentially classical non-symmetric one. The problem is that a detailed understanding of the process that leads, in the absence of observers or measurements, from one to the other, is lacking, rendering the standard account unsatisfactory (in fact, it is easy to show that the standard quantum evolution, via Schrödinger equation, cannot account for the breakdown of the initial symmetry).\footnote{It is worth mentioning that J. Hartle long ago noted the serious difficulties faced in attempting to apply quantum theory to cosmology, \cite{Hartle2005ie}. This lead him and his collaborators to conclude that some modified version of quantum theory was required. They turned to the Consistent Histories framework, about which we will say more later.}  The spontaneous reductions of objective collapse models, on the other hand, provide an explicit, observer-independent, mechanism for transitions form symmetric no non-symmetric states to occur. This feature was used in \cite{Origin} to address the problem in an inflationary cosmological context. Moreover, this application might not only account for the origin of the seeds of cosmic structure, but also may provide, through comparison with CMB data, with valuable clues for the construction of successful objective collapse models (see e.g., \cite{CC1,CC2,CC3}). 
 
\item The so-called problem of time in quantum gravity emerges from the broad disparity between the way the concept of time is used in quantum theory and the role it plays in general relativity. As a result, at least according to an important class of theories, the ``wave function of the universe'' does not seem to depend on time, rendering time inexistent at a fundamental level. Application of objective collapse models to quantum gravity, however, may dissolve the problem by providing objective means to anchor time fundamentally. That is because, in such theories, time evolution is governed by a modified equation that produces changes even if the Hamiltonian does not do so in the standard picture.
 
\item The black hole information paradox arises from an apparent conflict between Hawking’s black hole radiation and the fact that time evolution in quantum mechanics is unitary. The problem is that while the former suggests that information of a system falling into a black hole disappears (because, independently of the initial state, the final one will be thermal as a result of the Hawking evaporation), the latter implies that information must be conserved (because such evolution can be encoded in a unitary matrix which is necessarily invertible). There is, in fact, an ongoing, prominent debate regarding this paradox (there is not even agreement on whether the situation truly represents a paradox). The disagreements involve, among others issues, different positions regarding the singularity inside the black hole (whether it can be seen as destroying, or even encoding, the missing information) or whether quantum gravity will resolve the singularity and its effect on the missing information. It is evident, however, that the issue crucially depends on taking quantum theory to be fully information preserving. Therefore, if the fundamental quantum theory is taken to involve a degree of information destruction/creation (as objective collapse models do) the conflict, at least in principle, disappears. The critical issue, of course, is whether it is possible to solve the problem not only qualitatively but also quantitatively. We have proposed that this is indeed achievable by making the degree of departure from the Schr\"odinger equation dependent on the local value of the Weyl curvature (a choice that, moreover, seems to connect nicely with Penrose's famous “Weyl curvature hypothesis”). Therefore, by adopting an objective collapse model with these characteristics, the paradox seems to simply evaporate. 
\end{itemize}

Inspired by the last point, in \cite{Oko.Sud:14} we have also put forward a speculative idea connecting the spontaneous collapse events of objective collapse theories with black holes. Perhaps, we thought, the lack of unitarity of such theories is simply a reflection of the effects of virtual black holes that are created and destroyed in association with quantum fluctuations of the space-time metric. Maybe, then, ordinary quantum theory is what remains from the fundamentally time-irreversible and information destroying quantum world in situations where the effects of virtual black holes can be ignored. Our proposal seems to match well with the old idea that the laws of black hole mechanics imply a deep connection between quantum theory, relativity and thermodynamics, as well with early discussions by Penrose, Hawking and others in this general direction (see e.g., \cite{Pen,Haw}). Needless to say, much work is required before this scenario could be regarded as an acceptable description of nature.

%%%%%%%%%%%%%%%%%%%%%%%%%%%%%
\subsection{Crull's analysis of our work}
%%%%%%%%%%%%%%%%%%%%%%%%%%%%%
Finally we will examine what Crull has to say regarding the application of objective collapse models to the three problems mentioned above and, as she says, what decoherence reveals in each of those situations. Of course, much of her analysis depends on the false premise that decoherence solves the measurement problem, but that is not the end of her confusion. Below we will expose the many limitations in her evaluation of our work. We will start with the issue of the seeds of cosmic structure.
%%%%%%%%%%%%%%%%%%%%%%%%%%%%%
\subsubsection{The seeds of cosmic structure}
%%%%%%%%%%%%%%%%%%%%%%%%%%%%%
To start the discussion regarding the seeds of cosmic structure, Crull does little else than repeating 
%(a particularly poor version of) 
the standard story of how quantum fluctuations are enough to solve the problem:
\begin{quotation}
\noindent [I]n a fundamentally quantum universe there is no true vacuum, as even fields in vacuum states undergo quantum fluctuations. It turns out that these fluctuations, when applied to the inflaton field, are sufficient to give rise to the variety of structure now observable; hence, quantum fluctuations are the seeds of cosmic structure. \cite[p. 1038]{Crull}
\end{quotation}
The first line is enough to recognize her profound confusion. The vacuum state in standard quantum settings, such as a quantum field on a stationary spacetime (i.e., one with a global time-like Killing filed) is a technical notion \emph{defined} to be the state with the lowest possible energy. In non-stationary situations, such as an expanding universe, there is no well-defined notion of energy, and one has to face the fact that there are multiple constructions of the quantum field theory, leading to various possible vacuum states.\footnote{Things get further complicated by the fact that these constructions turn out to be inequivalent. However, a careful analysis using the algebraic approach shows that these problems can be readily overcome \cite{Wald-QFTin CS}.} In the cosmological scenario under consideration, the adequate state to choose is the so-called Bunch-Davies vacuum, which is the state that, in the asymptotic past, when the expansion rate was negligible, corresponded to the Minkowski vacuum.\footnote{Strictly speaking, if the expansion of the universe is not exactly exponential, and the space-time is therefore not truly described by the de Sitter line element, the state is not the Bunch-Davies vacuum. However, the important point for our purposes is that in such scenario the vacuum is still homogeneous and isotropic.} So, technically speaking, in an inflationary quantum universe, the vacuum state is a well-defined state, which is in fact the state used in computing, within the standard approach, the so-called power spectrum. The fact we highlight in \cite{Oko.Sud:14} is that such state is homogeneous and isotropic. Then, why is it, according to Crull, that in a fundamentally quantum universe there is no true vacuum? She does not seem to be referring to the issue of non-stationarity, but to the quantum aspects of the treatment.

The second part of her statement clarifies it for us. She argues that there is no true vacuum because ``the state undergoes quantum fluctuations.'' But this is not a tenable position because it implies, for example, that in the case of a simple harmonic oscillator there is no ground state because such state involves fluctuations in position and momentum. In fact, these so-called fluctuations are nothing else but the quantum uncertainties in the position and momentum operators evaluated on the ground state. This type of confusions are tied to an imprecise and often colloquial use of the word fluctuation. The truth is that the notion of a quantum fluctuation is often employed in conjunction of vague ideas associated with the developmental stages of quantum theory and is often used to hide poorly understood conceptual issues in modern quantum theory. For example, the vacuum state in a standard QFT is often said to undergo quantum fluctuations (see the last quote) and some even imagine these fluctuations as occurring in time or as statistical fluctuations in some ill defined ensemble. However, it is clear that the ground state of, say, a harmonic oscillator, even if it has fluctuations (i.e., uncertainties) in momentum and position, being an energy eigenstate, does not evolve in time. So which one is it, does it fluctuate in the usual meaning, as something changing rapidly and stochastically with time, or not? Of course it does not. So one has to take with a grain of salt any explanation given in terms of quantum fluctuations.

By looking at the often muddled discussions about these issues in some detail, one sees that the word ``fluctuation'' is frequently inadvertently used with different meanings, leading to confusion. In particular, in search for clarity, one must be careful not to confuse quantum uncertainties with statistical fluctuations. That is, we should distinguish between the variance of a certain quantity in an ensemble of systems and the quantum uncertainty of the corresponding quantity in the quantum state describing a single system. When Crull says ``these fluctuations, when applied to the inflaton field, are sufficient to give rise to the variety of structure now observable; hence, quantum fluctuations are the seeds of cosmic structure'' she is clearly failing to see the difference. Crull only rehearses the standard account of how \emph{quantum fluctuations} are supposed to give rise to the seeds of cosmic structure. However, she just repeats these words without explaining how the process is supposed to work. That is, she does not, nor anybody else for that matter, give a precise account of how, with or without ``quantum fluctuations'' (which, as we have said, should be better referred to as ``quantum uncertainties''), the symmetry of the initial state could be broken. 

At any rate, the key argument against the standard explanation for the emergence of seeds of cosmic structure, particularly in the context considered by Crull in which \emph{closed} quantum systems \emph{always} evolve according to the Schr\"odinger equation, is the following:
\begin{description}
\item[i)] The initial state for the whole universe is totally homogeneous and isotropic.
\item[ii)] The time evolution of such state, given that the universe is a closed system, is always controlled by a purely unitary dynamics.
\item[iii)] Such unitary dynamics preserves the homogeneity and isotropy of the initial state.
\item[iv)] As a result, regardless of possible interactions between different parts of the system, the state of the whole universe, at any time, is necessarily homogeneous and isotropic.
\end{description}
Therefore, regardless of the image that the words ``quantum fluctuations'' can bring to one's mind, 
 and in spite of the fact that different parts of the system may interact and get correlated, the standard story (even including decoherence) is incapable of explaining the emergence of structure out of the initially homogeneous and isotropic state. That is, no matter how complicated the internal dynamics may be, the basic assumptions of the standard story guarantee that the state will always be symmetric.\footnote{The simplicity of the structure of the previous argument can be illustrated with the following straightforward example: Suppose that we have a classical system, as complicated as you like, but such that, at $t=0$, its total energy is zero. Suppose, moreover, that the Hamiltonian of the system is time-translation invariant. As a result, the total energy of the system, at any other time, and independently of the details of the evolution, will also be zero. The same is true of the symmetry of the Bunch-Davies state under standard evolution.} 
 
Regarding this extremely simple argument, Crull writes:
\begin{quotation}
\noindent What is wrong with this conclusion is that it fails to account for inevitable non-local quantum interactions when assuming that the initially symmetric state describes a closed system. Even were the system to begin in a pure state, it would not remain so for long; along comes decoherence, and things are not what they seem. \cite[p. 1039]{Crull}
\end{quotation}
First, note her strange use of the notion of ``non-local quantum interactions.'' All interactions in standard cosmology are of course local. Presumably what Crull has in mind are non-local \emph{correlations} within the state but such correlations, as we explained above, have nothing to do with the breakdown of the symmetry. It seems, however, that she realizes that the complicated interactions by themselves cannot account for the breakdown of the symmetry thus she seeks help in decoherence. The idea seems to be that somehow, due to decoherence, an initially pure state for the universe as a whole will soon stop being so. Clearly this is absurd. Even if decoherence explains why states of subsystems lose purity, this would not apply to the state of the universe as a whole because there is nothing external to the universe, which could interact and get entangled with it.
 
Regardless, Crull then argues that ``quantum dynamics alone explains that such symmetries are not in fact destroyed but only become hidden'' and she points out that ``[t]he only measurement required to explain evolution from symmetry to apparent asymmetry is a (likely arbitrary) interaction with some external degree(s) of freedom'' \cite[p. 1039]{Crull}. We wonder what are supposed to be these degrees of freedom which are supposed to be external to the universe as a whole. Probably what Crull has in mind is not external degrees of freedom but internal ones that are \emph{ignored} because we are not interested on them or even because we lack the ability to keep track of them (either due to technical limitations or even for some stronger reasons, e.g., the degrees of freedom might lie beyond our cosmological horizon). However, such position is untenable; let us see why. 

According to standard inflationary cosmology, the quantum fluctuations of the inflaton field, which break the symmetries of homogeneity and isotropy of the vacuum state, are the starting point of the evolution of structures of the universe. That is, after such breakdown, the regions that turn out to have slightly larger densities than average, due to the attractive nature of the gravitational interaction, are supposed to become those region where matter accretes, leading eventually to the formation of galaxies, stars, planets, and eventually, in at least one of those, to the emergence of life. Life, of course, then is supposed to evolve according to the basic scheme proposed by Darwin and eventually, in one of the continents of that planet, a particular lineage of apes, takes an evolutionary path that leads to the development of relatively large brains and to what we call intelligence. These creatures then create civilizations and eventually invent science and discover quantum physics. Later, in contemplating their study of the universe, they, or at least some of them, decide to \emph{ignore} (which, as far as we know, is an essentially human action) certain degrees of freedom, or perhaps they simply find that they are not able to keep tack of them. By doing so, they obtain a decohered density matrix for the other degrees of freedom. This decoherence, we are told, is then supposed to be essential for an account of the breakdown of the symmetry that leads to all that preceded it. 

Clearly all this is very strange, to say the least. A human act is supposed to be, at least in part, the cause for the breakdown of the symmetry that leads to the emergence of galaxies, planets and, eventually, humans. That is, we are confronted with a clear case of closed causal explanation, for which the word circularity might be redundant but for which the word coherence certainly is inapplicable. It should be clear in any event that what humans decide or not to ignore cannot be part of a fundamental description of the world. 
 
Crull then goes on to describe the work of Kiefer et al. (see e.g., \cite{Kie}), where various other arguments are brought into play. This we can see as an implicit acknowledgment that what was offered before did not provide a sufficiently good argument. The point is, however, that a collection, no matter how large, of bad arguments, does not make up for a good one.

One of these additional arguments relies on squeezing, which arises as a feature of the evolution of some simple modification of a quantum mechanical harmonic oscillator. Such modification can be a simple as making the mass dependent on time. The point is that if the system was prepared in the ground state of the Hamiltonian at say, $ t=0$, then at all latter times the system would be in an excited state in which the uncertainties of the original canonical variables are much larger. The interesting aspect of all this is that there exist other canonical variables where the uncertainties will be again minimal. In any event, the problem is that such modified evolution does not break the initial symmetry and thus the state, squeezed or not, remains homogeneous and isotropic. In order to move forward, Kiefer et. al. rely on an unjustified interpretation of quantum theory. They claim that, because certain uncertainties are too large, one can consider the squeezed states as classical. This is not just contrary to the view taken by experimentalists working in quantum optics, who view these states as extremely quantum mechanical, but also clearly unwarranted. Moreover, the fact that the uncertainties are small when using a different set of variables, clearly renders the position inconsistent. 

A related argument, also put forward by Kiefer et. al., claims that, as a result of the squeezing, certain degrees of freedom become unobservably small and thus should be traced over. At such point the decoherence story is supposed to take over. However, it is clear that this would brings us back to a situation in which we explain our own existence in terms, at least in part, of our own limitations (for an in-depth discussion of the analysis presented by Keifer et. al. see \cite{Shortcommings}).

Finally, to put another nail in the coffin of these claims, we note that, in situations involving symmetries, such as the cosmological under discussion, the decoherence program is incapable of providing a preferential basis (i.e., the basis where the decohered matrix is diagonal). The issue can be seen in the EPR-B example presented in section \ref{ex}, where tracing over the spin degrees of freedom of one of the particles leads to a reduced density matrix for the other that is not only diagonal, but proportional to the identity. Therefore, it is diagonal in any orthogonal basis. That this situation would arise in the cosmological setting, follows from the following theorem first presented in \cite{Castagnino}:

\textbf{Theorem:}
Consider a quantum system made of a subsystem $S$ and an environment $E$, with corresponding Hilbert spaces $H_{S}$ and $H_{E}$ so that the complete system is described by states in the product Hilbert space $H_{S}\otimes H_{E}$. Let $G$ be a symmetry group acting on the Hilbert space of the full system in a way that does not mix the system and environment. That is, the unitary representation $O$ of $G$ on $H_{S}\otimes H_{E}$ is such that $\forall g \in G$, $\hat O(g) = \hat O^{S}(g)\otimes\hat O^{E}(g)$, where $\hat O^{S}(g)$ and $\hat O^{E}(g)$ act on $H_{S}$ and $H_{E}$ respectively. Let the system be characterized by a density matrix $\hat\rho$ which is invariant under $G$. Then the reduced density matrix of the subsystem is a multiple of the identity in each invariant subspace of $H_{S}$. 

Regarding the inflationary cosmology setting, this theorem indicates that, when offering an argument involving decoherence, even if one ignores all other shortcomings, when the selection of the degrees of freedom that will play the role of environment are made with any kind of objective criteria, (such as the argument that the modes of the scalar fields that as a result of inflation have a decreasing amplitude, and should thus be considered unobservable), the resulting reduced density matrix will not offer a unique selection of the preferred basis. This is simply because such matrix will be proportional to the identity in very large subspaces of the full Hilbert space of the theory. 

%%%%%%%%%%%%%%%%%%%%%%%%%%%%%
\subsubsection{The problem of time in quantum gravity}
%%%%%%%%%%%%%%%%%%%%%%%%%%%%%
Regarding the problem of time in quantum gravity, Crull says the following:
\begin{quotation}
\noindent The diffeomorphism invariance of general relativity, when carried into a quantized theory, results in a quantum state that cannot differentiate between space-times -- i.e., the quantum description fails to pick out a unique hyper-surface corresponding to our physical universe. \cite[p. 1041]{Crull}
\end{quotation}
It is hard to decide what to make of the above explanations. For starters, it mixes up the comparison of different space-times with the comparison of different hyper-surfaces in one space-time. The actual problem is that, in theories involving gauge symmetries (i.e., symmetries that are associated with multiple representations of the same physical situation), the wave function must assign the same value to each representation. This is the basic conceptual foundation of Dirac's quantization procedure for such systems. The problem is that in theories respecting general diffeomorphism invariance, such as general relativity, the data on any Cauchy hyper-surface is equivalent to the data on any other one corresponding to the same space-time. Thus, the wave function cannot depend on the choice of hyper-surface, and this leads to time disappearing from the theory. 

At any rate, commenting on our proposal, Crull states:
\begin{quotation}
\noindent [Okon and Sudarsky]’s proposal to resolve the problem of time by simply introducing nonunitary terms is motivated by the false impression that the dynamics of the system under consideration is properly unitary -- and we know this is hardly ever true. \cite[p. 1041]{Crull}
\end{quotation}
and then:
\begin{quotation}
\noindent [C]anonical approaches to quantum gravity have largely only considered “pure”, matterless gravity fields, taking for granted the pedestrian fact that any system short of the universe [at] large is in truth interacting and hence not evolving strictly in accordance with Schr\"odinger’s equation. \cite[p. 1042]{Crull}
\end{quotation}
Apparently, Crull believes that when a system is interacting, it does not evolve in accordance 
with Schr\"odinger’s equation. Perhaps what she is considering is the effective dynamics for just part of the quantum system. The confusion can be seen as arising from the belief that decoherence brings about a \emph{fundamental} breakdown of unitarity, in contrast with the simple, but rather unhelpful, fact that the \emph{effective} dynamics for a part of the system might not be unitary. This of course does not help in the context of interest because, at the fundamental level, one does not want to leave anything outside the quantum mechanical treatment since the system in question is the whole universe. In fact, as bringing gravity under the quantum umbrella would close the program of providing a quantum treatment of every fundamental degree of freedom in nature, deciding to leave something outside would be, in principle, unwarranted. Moreover, it is hard to consider situations where gravity is treated quantum mechanically and other fields are not. 

In sum, while decoherence might explain why \emph{subsystems} may effectively undergo non-unitary evolution, it could never imply that closed systems, such as the universe as a whole, would ever deviate from unitary evolution. We conclude that decoherence by itself is unable to help with the problem of time in quantum gravity. %In fact, as we will see below, even the strongest advocates for decoherence playing a fundamental role in these contexts, realize the need to rely on some extra input, such as an Everettian interpretation.

%%%%%%%%%%%%%%%%%%%%%%%%%%%%%
\subsubsection{The black hole information loss paradox}
%%%%%%%%%%%%%%%%%%%%%%%%%%%%%
Regarding the information loss paradox, Crull starts the discussion by acknowledging that the non-unitary behavior of the objective collapse models does resolve the issue. However, she argues that the problem can also be solved without having to assume non-unitary behavior at the fundamental level. In this respect, she states the following:
\begin{quotation}
\noindent [O]ne need not forsake unitary evolution, only the assumption of unitary evolution for systems partaking in Hawking radiation. This is the tactic described in [\cite{Gam1} and \cite{Gam2} by Gambini et al.]: the physical clocks picked out by decoherence are used to calculate the rate of information loss in black hole evaporation. \cite[p. 1042]{Crull}
\end{quotation}
Therefore, we are led again to consider a scheme where the underlying evolution is the standard unitary one, provided by the Schr\"odinger equation. However, it is argued that this refers only to a fundamental underlying time to which we have no access. Instead, we are directed to regard the evolution as described by an empirically accessible time, associated with a physical clock, the march of which, in terms of the fundamental time, is again ruled by its own Hamiltonian. The point made in \cite{Gam1} and \cite{Gam2} is that, when considering the evolution of the rest of the system, relative to the time measured by the physical clock, one does not recover an exact unitary evolution. This is then consider as a possible explanation for the apparent breakdown of unitarity in the complete evaporation of black holes. 

Does this provide a satisfactory resolution of the paradox? At first sight it might seem that it does. However, a moment's consideration reveals the deep flaw in the argument. Remember that, in terms of the fundamental time, the evolution would still be fully unitary. Therefore, what has been achieved, is to deviate attention from the truly fundamental question: where does the fundamental information reside, if in terms of the fundamental time the black hole eventually evaporates (even if the fundamental time for this is different from that indicated by the physical clock)? Instead, we have been led to consider a very different question regarding the description of things in terms of the physical clock. 

As explained in detail in \cite{The-montevideans}, the so-called ``Montevideo Interpretation'' described in \cite{Gam1} and \cite{Gam2}, in which the measurement problem is to be solved by relying on the decoherence brought about by the intrinsic limitations of physical clocks, should be considered in a sort of Everettian scheme, in which the full information is invariably present in the complete state of the quantum system. The point is that when considering the situation at a time well after the evaporation of the black hole, and independently of how precisely such time might be specified by real physical clocks, one cannot see, in any such proposal, where the information is supposed to be preserved, i.e., what are the degrees of freedom where such information would be encoded. 
 
In other words, the puzzle, when formulated in terms of the fundamental physical laws, is still present, and it only seems to disappear when we look at things in a less precise way. This seems to be in fact the general strategy of some advocates of the decoherence program when used to address the conceptual difficulties of quantum theory: to replace a deep real problem by a secondary one and to convince us not only that the latter is solved, but that doing so is equivalent to solving the former. No wonder that we feel like spectators in a magic show. The problem is that the tactic used by the advocates of such an approach is not very different than the one used by the magician. %Others of course acknowledge that decoherence by itself is hardly enough and, try to deal with its shortcomings with the reliance on an Everettian or Many Worlds interpretation.

%%%%%%%%%%%%%%%%%%%%%%%%%%%%%%%%%%%%%%%%%%%%%%%%%%%%%%%%%%%%%%
\section{Conclusions}
\label{Con}
%%%%%%%%%%%%%%%%%%%%%%%%%%%%%%%%%%%%%%%%%%%%%%%%%%%%%%%%%%%%%%

Attempts to use decoherence by itself to resolve foundational questions in quantum theory, such as the measurement problem, suffer from the same basic shortcoming confronted by those who wish to, as they say, ``have their cake and eat it too''. These attempts try to avoid modifications of quantum theory, but at the same time, want to obtain the benefits that such modifications can bring. In practice, people that adopt this attitude often end up adapting their interpretation of the theory in a case by case manner. Then, they try to justify such moves using arguments that rely on a combination of classical intuition and quantum mechanical elements, borrowed freely from either the old or the modern versions, without any concern for the fact that some of those arguments rest on assuming the very aspect they want to address. 

On the other hand, even some of the strongest advocates of the decoherence program acknowledge that decoherence by itself is hardly enough to solve foundational problems and realize the need to rely on some extra input (see, e.g., \cite{Zura}). As a result, some are driven to look for the missing aspects by complementing decoherence with interpretational elements from, e.g., Many Worlds \cite{Wal} or the Consistent Histories formalism \cite{CH1,CH2}. However, we do not believe that these approaches can at present be considered satisfactory. Regarding the Consistent Histories approach, we point the interested reader to the critical works \cite{aCH1,aCH2,aCH3}. As for the Many Worlds Interpretation, we can point to section 4 of \cite{MW}. At any rate, what is clear is that, without a clear interpretational framework, one is at a loss not only regarding the exceptional situations we have dealt with here in some detail, but also with respect to standard applications of the theory.

We find rather puzzling the relatively widespread willingness to accept analyses of foundational issues in quantum mechanics, which seem to work when advanced with an imprecise language, but that clearly fail dramatically when carried out in a rigorous and detailed manner. One cannot help but recall the impetus behind efforts to design perpetual motion machines, using complex contraptions of wheels, pulleys and levers, in the unexplainable and stubborn hope of somehow bypassing the second law of thermodynamics.

Perhaps, just as the desire to avoid confronting the inevitability of death predisposes people to accept rather fantastic stories, which they would not even consider in a different context, the desperate desire to avoid confronting the difficulties of quantum theory allows people to be deceived or even to deceive themselves. This is certainly understandable as a human psychological trait, but as far as the goal to achieve a deeper and better understanding of nature is concerned, it certainly is an impediment. On the contrary, what can be of help is a disciplined and unflinching commitment to maintain coherence in our theoretical and philosophical analyses. Carefully assessing the extent to which a proposal might work is the only path to advancement in a field where there are not too many empirical clues. It is far more productive to consider in detail clear ideas that might be wrong, that to entertain unclear and vague arguments in ways that might even be self-contradictory. As noted by Sir Francis Bacon when considering the scientific enterprise in general: ``Truth emerges more readily from error than from confusion.''

%%%%%%%%%%%%%%%%%%%%%%%%%%%%%%%%%%%%%%%%%%%%%%%%%%%%%%%%%%%%%%
\section*{Acknowledgments}
%%%%%%%%%%%%%%%%%%%%%%%%%%%%%%%%%%%%%%%%%%%%%%%%%%%%%%%%%%%%%%
We acknowledge partial financial support from DGAPA-UNAM project IG100316.
%%%%%%%%%%%%%%%%%%%%%%%%%%%%%%%%%%%%%%%%%%%%%%%%%%%%%%%%%%%%%%
%%%%%%%%%%%%%%%%%%%%%%%%%%%%%%%%%%%%%%%%%%%%%%%%%%%%%%%%%%%%%%
%%%%%%%%%%%%%%%%%%%%%%%%%%%%%%%%%%%%%%%%%%%%%%%%%%%%%%%%%%%%%%
%\bibliographystyle{ieeetr}
%\bibliography{biblioCH.bib}
%%%%%%%%%%%%%%%%%%%%%%%%%%%%%%%%%%%%%%%%%%%%%%%%%%%%%%%%%%%%%%

\end{document}